\documentstyle[aps,amssymb,epsfig,multicol]{revtex}
\def\be{\begin{equation}}
\def\ee{\end{equation}}
\def\bq{\begin{eqnarray}}
\def\eq{\end{eqnarray}}
\def\bm{\begin{multicols}{2}}
\def\em{\end{multicols}}
\def\Rc{\check{R}}
 1

\begin{document}

\draft
\title{Recent results on Integrable electronic models}

\author{Alberto Anfossi${}^1$, Fabrizio Dolcini${}^2$ and Arianna Montorsi ${}^1$ \\
${}^1$ Dipartimento di Fisica and Unit\`a INFM, Politecnico di
Torino, I-10129 Torino, Italy \\
${}^2$ Physikalisches Institut, Albert-Ludwigs-Universit\"at,
79104 Freiburg, Germany}

\date{2 December 2003}
\maketitle
\begin{abstract}
We review the approach of generalized permutator to produce a
class of integrable quantum Hamiltonians, as well as the technique
of Sutherland species (SS) to map a subclass of it into solvable
spinless fermions models. In particular, we apply the above scheme
to construct integrable interacting electron Hamiltonians: first
we review the extended Hubbard case, discussing both ground state
and thermodynamics; then we pass to constrained fermion models,
generating 56 integrable cases, among which both supersymmetric
$t-J$ model and infinite $U$ Hubbard model are obtained, as well
as other physically interesting cases, such as a particular $t-V$
model. For the latter we describe how the complete spectrum can be
gained by means of SS technique. Finally we speculate about
possible applications to spin $S$ models. 
\end{abstract}

\begin{multicols}{2}
\section{Introduction}
In recent years the discovery of materials that exhibit, at least
in some energy regimes, a strong one-dimensional character has
renewed the interest in models of interacting electrons in low
dimensional lattices.
\\Conducting polymers such as polyacetylene
$\mbox{(CH)}_x$ was probably the first noticeable example of such
type of materials\cite{1Dbook}, where the interplay between
$\pi$-electron
 correlation and chain dimerization was intensively
studied\cite{TINKA}.
\\Successively, a very rich
temperature-pressure phase diagram has been observed for Bechgaard
salts\cite{BECH}, which are linear-chain organic compounds such as
$\mbox{(TMTSF)}_2 \mbox{X}$ and $\mbox{(TMTTF)}_2 \mbox{X}$, where
$\mbox{ClO}_4$ or $\mbox{X}=\mbox{Br}$. In considerable part of
their phase diagram, namely at sufficiently high temperatures and
pressure, the inter-chain hopping is irrelevant and such materials
actually behave as 1D systems.
\\More recently, huge interest has been devoted to carbon
nanotubes\cite{SWCN}, which can be regarded to as graphite
cylinders of a typical radius of some nm. Due to the discrete
nature of the radial motion, only one band is actually involved in
determining the  electronic properties (at least up to 1eV energy
scales); for this reasons, as well as for the fact that they are
ballistic, such materials are nowadays considered as the most
promising material to probe electronic correlations in 1D.
\\Further significant example are semiconductor-based quantum
wires synthesized through cleaved edge overgrowth\cite{semi-QW},
as well as Edge states in Fractional Quantum Hall Effect systems.\\

In modelling any of the above materials, it is crucial to take
into account the electron-electron interaction, since the latter
heavily determines their physical behavior. It is indeed well
known that in low dimension {\it correlation} effects become
important, even for moderate interaction strength; this boils down
to the general fact that in 1D a single particle picture is not
valid. The conventional techniques that are used in 3D to treat
many body systems are therefore either unreliable (such as mean
field) or inapplicable (such as Fermi Liquid theory), when
investigating 1D materials. For this reason, alternative
approaches have been formulated and developed, both analytical
(e.g. Luttinger Liquid
Theory\cite{LL}) and numerical(e.g.DMRG\cite{DMRG}).\\

In this general context, the availability of exact results for 1D
electronic models (at least for some values of the model
parameters) plays a crucial role, not only as a conceptual
outcome, but also as a comparison test for the above
'non-traditional' techniques.\\

Within the exact-result approaches to one dimensional systems, the
{\it Coordinate Bethe Ansatz} (CBA) is probably the most famous
technique; it amounts to prove, when possible, that a given model
Hamiltonian has eigenfunctions of the form proposed by
Bethe\cite{BETHE}; this method has been quite successfully applied
to models of correlated electrons. For instance CBA has been used
for solving the Hubbard model\cite{HUB} and the t-J
model\cite{tJ}, which are somehow the prototype models to account
for correlation effects in electron systems. The former reads
\begin{eqnarray}
{\mathcal H}_{\rm Hub} &=& - t \sum_{j=1}^L
\sum_{\sigma=\uparrow,\downarrow} ( c^{\dagger}_{j,\sigma}
c^{}_{j+1, \sigma} +c^{\dagger}_{j+1,\sigma} c^{}_{j+1, \sigma}) +
 \label{HUBHAM}   \\
& &  \hspace{2cm} + U \sum_{j=1}^L
\hat{n}_{{j},\uparrow}\hat{n}_{{j}, \downarrow} \nonumber
\end{eqnarray}
where  $c^\dagger_{j,\sigma}$ and $c^{}_{j, \sigma}$ are  electron
creation and annihilation operators on a 1D lattice, $j$ and
$\sigma$ label position and spin of the electons on the chain, and
$\hat{n}_{{j}, \sigma}=c^\dagger_{j,\sigma} c^{}_{j, \sigma}$ is
the number operator. In Eqs.(\ref{HUBHAM}) the first term is the
tight-binding part and the second the on-site Coulomb repulsion
competing with it. The exact solution states that any small $U>0$
causes the ground state of the system to be insulating at
half-filling, confirming the dramatic effect that correlations
have in 1D.
\\The t-J model involves a spin-spin coupling term and its Hamiltonian reads
\begin{eqnarray}
{\cal H}_{\rm t-J} &=& -t \sum_{j=1}^L
 (1-\hat{n}_{j, \bar{\sigma}})
(c^{\dagger}_{j,\sigma} c^{}_{j+1,\sigma} +h.c.) (1-\hat{n}_{j+1,
\bar{\sigma}})
 +  \nonumber \\
& & \hspace{1cm} + J \sum_{j=1}^L (\vec{{\mathcal{S}}}_{j} \cdot
\vec{{\mathcal{S}}}_{j+1} -\frac{1}{4} \hat{n}_{j} \hat{n}_{j+1} )
\label{tJHAM}
\end{eqnarray}
where $\vec{\mathcal{S}}=c^\dagger_{j,\sigma}
\frac{\vec{\sigma}_{\sigma \sigma'}}{2} c^{}_{j,\sigma'}$ is the
spin operator, and $\bar{\sigma}=\downarrow$ (resp.$\uparrow$) if
$\sigma=\uparrow$ (resp.$\downarrow$). The exact solution of the
t-J model for $J=2 t$ was the first analytical proof that the
Luttinger liquid hypothesis for 1-D systems was well grounded.
Indeed the ground state of this model, for any filling value, is a
liquid of singlet bound pairs of varying spatial separation and
binding energy. The excitations were actually found to be of two
types: spin-like and charge-like, and they can be proved to be
gapless. \\The t-J model is also an example of constrained
fermions, in the sense that the double occupied states are a
priori excluded from the Hilbert space; this type of constraint
physically emerges e.g. when investigating the low-energy singlet
excitations of multi-band models; in the regime of strong
correlation an effective 1-band Hamiltonian can be
obtained\cite{Zhang-Rice,EMERY}.
\\A quite detailed review of CBA-solved models
can be found in\cite{EKSbook}.\\

Beyond the CBA, there exist another quite useful technique within
the framework of exact-result approaches. It is called {\it
Quantum Inverse Scattering Method} (QISM)\cite{QISM}, and is in a
sense more general than the CBA, in that it directly involves the
notion of {\it integrability} of the Hamiltonian $\mathcal{H}$.
Here integrability actually consists in the possibility of finding
a complete set of mutually commuting observables
\begin{equation}
[ {\mathcal J}_{n} \, , \, {\mathcal J}_{n'}] \, = \, 0 \, ,
\end{equation}
so that the eigenstates of $\mathcal{H}$ can be univoquely
characterized by the quantum numbers related to such observables.
In the QISM, all the conserved quantities are encoded in an
equation, known as the Yang-Baxter-Equation (YBE); each solution
of the YBE determines a specific model, endowed with a set of
conserved quantities. The standard way to determine the
eigenstates within the QISM is again based on an Ansatz, called
Algebraic Bethe Ansatz (ABA); differently from the CBA, the ABA
explicitly lies on the underlying symmetries of the
Hamiltonian contained in the related YBE.\\

Within the above scenario, in this paper we first briefly
summarize in sec.\ref{secII} the main ideas of the QISM applied to
fermionic models; then, in sec.\ref{secIII} we review a recently
developed method\cite{DOMO1} to determine integrable electron
models. This method, formulated within the QISM, amounts to
finding solutions of YBE of {\it polynomial} form. We show in
sec.\ref{secIV} that already the easiest case of first order
polynomial yields a rich variety of integrable
models\cite{DOMO_SGE}. In section \ref{secV} we describe a
technique (different from the usual ABA) to obtain the complete
spectrum of the Hamiltonian for a subclass of the integrable
cases. In particular, in sec.\ref{secVI} we review the integrable
cases for extended Hubbard models (dimension of the local vector
space $d_{V}=4$), whereas in sec.\ref{secVII} we explicitly derive
the integrable cases for $d_{V}=3$, and look at their
implementation by means of constrained fermions. Finally, in sec.
\ref{secVIII} we give some conclusions.\\
\section{Integrability for electronic models}
\label{secII}
\subsection{Some introductory  material}
\label{sec_IIa} In the second quantization formulation, a system
of fermions on a 1D lattice is described by creation and
annihilation operators, which are governed by the algebra:
\begin{equation}
\{ c^{}_{{i},s}, c^{}_{{j}, s'}\} = 0 \hspace{1cm} \{ c^{}_{{i},s}
, c^{\dagger}_{{j}, s'} \} = \delta_{{i},{j}}\, \delta_{s,s'}
\label{antic}
\end{equation}
where $s=(a,\sigma)$ is in general a multi-label, accounting for
the orbital (=$a$) and for the spin (=$\sigma$). Here $a$ can vary
from 1 to $N_{\rm orb}$ and $\sigma$ can take $2J+1$ values, where
$N_{\rm orb}$ is the number of orbitals and $J$ the (half-integer)
spin. \\

In lattice model, one can associate to each site a local vector
space $V_j$. As an example, in the case of $s$-orbital electrons
we have $a=s$ and $\sigma=\uparrow, \downarrow$, $V_j$ is made up
of the 4 vectors
\begin{eqnarray}
V_j &=& {\mbox{\it Span}} \left( \, | \uparrow \rangle_j =
c^{\dagger}_{j\,\uparrow} |o \rangle_j \, | \downarrow \rangle_j
=c^{\dagger}_{j\,} |o \rangle_j \, ; \right. \nonumber
\\
& & \left. \,\hspace{1cm} |o \rangle_j \, ; \, | \downarrow
\uparrow \rangle_j =c^{\dagger}_{j\,\downarrow}
c^{\dagger}_{j\,\uparrow} |o \rangle_j \right)
 \label{vectors-s}
\end{eqnarray}
where $| o \rangle_j$ is the local vacuum.
\\More generally, $V_j$
has dimension $d_V$ and is spanned by vectors $| \alpha \rangle_j$
\begin{equation}
V_j = {\mbox{\it Span}} \left( \, | \alpha \rangle_j =
h^{(\alpha)}_j |o \rangle_j \, , \alpha=1 \ldots d_V \right)
\label{vectors}
\end{equation}
Here the $h^{(\alpha)}_j $'s are products of creation operators
$c^{\dagger}_{j s}$ with different $s$ \cite{nota1}. \\

An important property of fermionic systems is that, due to the
anti-commutation relations (\ref{antic}), the space $V_j$ has an
intrinsic {\it graduation}; this means that $V_j$ can be
decomposed into an odd and an even subspaces ($V_j=V^{(1)}_j
\oplus V^{(0)}_j$), where the odd (even) subspace $V^{(1)}_j$
($\,V^{(0)}_j$) is spanned by those vectors that are built with an
odd (even) number of creation operators~$c^{\dagger}_{i s}$. For
instance, in the case of the $s$-orbital electrons, the subspace
spanned by $|\uparrow\rangle_j$ and $|\downarrow\rangle_j$ is odd,
and the subspace spanned by $|o \rangle_j$ and
$|\downarrow\uparrow\rangle_j$ is even.
  Similarly, the space  of local
linear operators acting on $V_j$ (denoted by End($V_j$)) is also
graded; odd and even vectors and operators are also said to be
parity- homogeneous; in particular they are  respectively said to
have parity $p=1$ and $p=0$, so that for any homogeneous
${\mathcal{O}}^{(a)}_j , {\mathcal{O}}^{(b)}_j \, \in
\mbox{End}(V_j)$ the relation $p({\mathcal{O}}^{(a)}_j \,
{\mathcal{O}}^{(b)}_j)=p({\mathcal{O}}^{(a)}_j)+
p({\mathcal{O}}^{(b)}_j)$ holds. Technically this is expressed by
saying that End($V$) is a graded local algebra.\\

Another important ingredient related to the vector space $V_j$,
which plays an important role in the QISM to be presented below,
is the state projectors. They are  defined as
\begin{equation}
{{\mathcal{E}}_{j}}^{\beta}_{\alpha}=|\alpha \rangle_j \,
{}_j\langle \beta| \label{projectors}
\end{equation}
The projectors fulfill very important properties:
\begin{eqnarray}
\left[ {{\mathcal{E}}_{j}}^{\beta}_{\alpha} \, , \,
{{\mathcal{E}}_{k}}^{\delta}_{\gamma} \right]_{\pm} = 0
\hspace{1cm} \forall j \neq k \label{prop-1}
\\
{{\mathcal{E}}_{j}}^{\beta}_{\alpha} \,
{{\mathcal{E}}_{j}}^{\delta}_{\gamma} = \delta^{\beta}_{\gamma} \,
{{\mathcal{E}}_{j}}^{\delta}_{\alpha} \hspace{2.2cm}
\label{prop-2}
\end{eqnarray}
where $[ X , Y]_\pm=X \, Y -(-1)^{p(X)p(Y)} \, Y X$. \\In the case
of $s$-orbitals  they are explicitly given by the entries ($a-$th
row and the $b-$th column) of the following $4 \times 4$ matrix
\end{multicols}
\begin{equation}
{\mathcal{E}}_{j} \, = \, \pmatrix{
n^{}_{j\uparrow}(1-n^{}_{j\downarrow}) & c^{\dagger}_{j\uparrow}
c^{}_{j\downarrow} & c^{\dagger}_{j\uparrow}
(1-n^{}_{j\downarrow}) & c^{}_{j\downarrow} n^{}_{j\uparrow} \cr
c^{\dagger}_{j\downarrow} c^{}_{j\uparrow} & n^{}_{j\downarrow}
(1-n^{}_{j\uparrow}) & c^{\dagger}_{j\downarrow}
(1-n^{}_{j\uparrow}) & -c^{\dagger}_{j\uparrow} n^{}_{j\downarrow}
\cr
c^{}_{j\uparrow} (1-n^{}_{j\downarrow}) & c^{}_{j\downarrow}
(1-n^{}_{j\uparrow}) & (1-n^{}_{j\uparrow})(1-n^{}_{j\downarrow})
& c^{}_{j\uparrow} c^{}_{j\downarrow} \cr
c^{\dagger}_{j\downarrow} n^{}_{j\uparrow} & -c^{}_{j\uparrow}
n^{}_{j\downarrow} & c^{\dagger}_{j\downarrow}
c^{\dagger}_{j\uparrow} & n^{}_{j\downarrow} n^{}_{j\uparrow} \cr
} \label{Hubb-proj}
\end{equation}
\begin{multicols}{2}
For constrained fermions the projector matrix is just the left
upper $3 \times 3$ sub-matrix of (\ref{Hubb-proj}). \\Notice that
each of the entries of the above matrix is an homogeneous operator
with parity $p({{\mathcal{E}}_j}^{\beta}_{\alpha}) \, = \,
p(\alpha)+p(\beta)$\cite{nota2}. Projectors are useful in
decomposing any operator into its fundamental state-processes: any
single-site operator ${\mathcal{O}}^{(a)}_j$ can indeed be written
as
\begin{equation}
{\mathcal{O}}^{(a)}_j=({O}^{(a)})^{\alpha}_{\beta} \,
{{\mathcal{E}}_{j}}^{\beta}_{\alpha} \label{local-local}
\end{equation}
where $({O}^{(a)})$ is its representing matrix, defined through
\begin{equation}
{\mathcal{O}}^{(a)}_j \, |\beta \rangle_j  =
({O^{(a)}})^{\alpha}_{\beta} \, |\alpha \rangle_j \label{local}
\end{equation}
\subsection{Quantum Inverse Scattering Method for fermions: a short and schematic review}
\label{IIb} The Quantum Inverse Scattering Method (QISM) is a
powerful tool for studying quantum integrability because it
provides models endowed with a set of mutually commuting
operators. Within the QISM a key role is played by the Yang-Baxter
Equation (YBE)
\begin{eqnarray}
& &({\bf I} \otimes \Rc(u-v)) \, (\Rc(u) \otimes {\bf I}) \,
({\bf I} \otimes \Rc(v)) \, =  \nonumber \\
& & \, \, \, =(\Rc(v) \otimes {\bf I}) \, ({\bf I} \otimes \Rc(u))
\, (\Rc(u-v) \otimes {\bf I}) \label{YBE}
\end{eqnarray}
Here $\Rc$ is a $d_V^2 \times d_{V}^2$  $C$-number matrix, and
${\bf I}$ is the $d_V \times d_{V}$ identity matrix ($d_V$ being
the dimension of the local vector space $V_j$); $\Rc$ depends on
the complex number $u$ (called spectral parameter), and
thus the YBE is a functional equation for $\Rc$.\\

To each solution of the YBE one can associate an integrable model;
here below we shall briefly describe how to obtain integrable {\it
fermionic} models (fermionic QISM, for more details see
Ref.\cite{GOMU,DOMO_NUCLPHYSB}).
\\From the $\Rc$-matrix  one can construct an operator-valued matrix ${\mathcal{L}}_j$,
related to each site $j$, as follows
\begin{equation}
{\mathcal{L}_j}^{\alpha}_{\beta}(u)=(-1)^{p(\alpha) p(\gamma)}
\Rc^{\gamma \alpha}_{\beta\delta}(u) \,
{{\mathcal{E}}_j}^{\delta}_{\gamma} \label{L-op}
\end{equation}
where the ${{\mathcal{E}}_j}^{\delta}_{\gamma}$ are the projectors
(\ref{projectors}). Here $\alpha$ gives the row and $\beta$ the
column of the entry of the ${\mathcal{L}}_j$ matrix (usually
called ${\mathcal{L}}$-operator). Due to eqn.(\ref{prop-1}) the
entries of two operators ${\mathcal{L}}_j$ and ${\mathcal{L}}_k$
of different sites fulfill
\begin{equation}
\left[{{\mathcal{L}}_j}^{\alpha}_{\beta}(u),{{\mathcal{L}}_k}^
{\gamma}_{\delta}(u')\right]_{\pm}=0 \hspace{0.3cm} \forall
\alpha,\beta,\gamma,\delta \hspace{0.5cm} \forall u,u'
\hspace{0.5cm} \forall j\neq k \label{L-op-sc}
\end{equation}
At the same time the property (\ref{prop-2}) of the projectors
allows to show that the Yang-Baxter equation (\ref{YBE}) is
actually equivalent to
\begin{eqnarray}
\lefteqn{\Rc(u-v) \, ({{\mathcal{L}}}_j(u) \otimes_{s}
{{\mathcal{L}}}_j(v)
)= } & & \label{RLLg} \\
& & \hspace{1cm} =({{\mathcal{L}}}_j(v) \otimes_{s}
{{\mathcal{L}}}_j(u) ) \, \Rc(u-v) \nonumber
\end{eqnarray}
where the symbol ${\otimes}^{}_{s} $ is called {\it graded} tensor
product, and it is defined as
\begin{equation}
\left({\mathcal{A}} {\otimes}^{}_{s} {\mathcal{B}}\right)^ {(\{
\alpha \},\{ \gamma \})}_{(\{ \beta \},\{ \delta \})} \, = \,
{\mathcal{A}}^{\{ \alpha \}}_{\{ \beta \}} \, {\mathcal{B}}^{\{
\gamma \}}_{\{ \delta \}} \, (-1)^{(p(\{\alpha\})+p(\{\beta\})) \,
p(\{\gamma\})} \label{tp_s}
\end{equation}
The presence of the above additional signs is strictly related to
the fermionic nature of the system, as observed in \ref{sec_IIa}.
\\Eq.(\ref{RLLg}) is a local relation, in that it only involves
the $j$-th site of the 1D lattice. However, using the property
\begin{equation}
({{\mathcal{L}}_i} {\otimes}^{}_{s} {{\mathcal{L}}_j} )
({{\mathcal{L}}_k} {\otimes}^{}_{s} {{\mathcal{L}}_l}) =
({{\mathcal{L}}_i} {{\mathcal{L}}_k} {\otimes}^{}_{s}
{{\mathcal{L}}_j}  {{\mathcal{L}}_l}) \hspace{1cm} \forall \, j
\neq k \label{prodotto-per-L}
\end{equation}
one can easily show that
\begin{eqnarray}
\lefteqn{\Rc(u-v) \, ({{\mathcal{T}}}(u) \otimes_{s}
{{\mathcal{T}}}(v) ) \,=}  & & \label{RTTg} \\
& & \hspace{1cm} = ({{\mathcal{T}}}(v) \otimes_{s}
{{\mathcal{T}}}(u) ) \, \Rc(u-v) , \nonumber
\end{eqnarray}
where
\begin{equation}
{{\mathcal{T}}}(u)={{\mathcal{L}}}_L(u) \ldots
{{\mathcal{L}}}_2(u) {{\mathcal{L}}}_1(u)
\end{equation}
Notice that ${{\mathcal{T}}}(u)$ is a operator-valued matrix
(usually called monodromy matrix), whose entries are now operators
defined on the whole 1D chain.
\\Eq.(\ref{RTTg}) is the main relation of the QISM, because
taking the trace of ${{\mathcal{T}}}(u)$ yields
\begin{equation}
\left[ tr \, {{\mathcal{T}}}(u) ,  tr \, {{\mathcal{T}}_N}(v)
\right] = 0 \hspace{1cm} \forall u,v \label{cons-tr-str}
\end{equation}
which already implies the existence of an infinite number of
conservation laws; indeed developing the operator $tr \,
{{\mathcal{T}}}(u)$ in powers of $u$, eq.(\ref{cons-tr-str})
states that all the coefficients (i.e. operators defined on the
chain) mutually commute. \\Typically one searches for solutions of
the YBE with the 'boundary condition' that for a given point
$u_0$, $\Rc(u)$ is the identity matrix
\begin{equation}
\Rc^{\alpha \gamma}_{\beta\delta}(u_0)=\delta^{\alpha}_{\beta} \,
\delta^{\gamma}_{\delta} \label{cond-2}
\end{equation}
In this case one typically introduces
\begin{equation}
{\mathcal{Z}}(u) := (tr \, {\mathcal{T}} (u_0))^{-1} \, tr \,
{\mathcal{T}} (u) \label{Z}
\end{equation}
and defines
\begin{equation}
{\mathcal{J}}_n \, = \, \left. \frac{d^n}{d u^n} \ln
{\mathcal{Z}}(u) \right|_{u=u_0} \hspace{1cm} n \ge 1 \label{J_n}
\quad .
\end{equation}
Eq.(\ref{cons-tr-str}) therefore implies
\begin{equation}
[{\mathcal{J}}_n \, , \, {\mathcal{J}}_{n'} \,] \, = 0
\end{equation}
where ${\mathcal{J}}_n$ are the mutually conserved quantities. It
can be shown that ${\mathcal{J}}_n$ is the sum of operators
involving clusters of no more than $n+1$ sites\cite{KUL}.
\\The first conserved quantity ${\mathcal{J}}_1$ is usually
interpreted as the Hamiltonian, and the other ones as its
symmetries. ${\mathcal{J}}_1$ has the form
\begin{equation}
{\mathcal{J}}_1 ={\mathcal{H}}=\sum_{j=1}^{L} {\mathcal{H}}_{j \,
j+1}
\end{equation}
with periodic boundary conditions ${\mathcal{H}}_{L \, L+1}=
{\mathcal{H}}_{L \,1}$. \\Each ${\mathcal{H}}_{j \, j+1}$ is a
two-site Hamiltonian, and it is straightforwardly connected with
the $\Rc$-matrix solution of the YBE. Explicitly one can show that
\begin{eqnarray}
\lefteqn{ {\mathcal{H}}_{j \, j+1} = }
\label{2-site-Ham-op} \\
& & =(-1)^{p(\gamma) (p(\beta)+p(\delta))} \left.
\partial_u \check{R}^{\alpha \beta}_{\gamma \delta}(u) \right|_{u=u_0}
{{\mathcal{E}}_j}^{\gamma}_{\alpha} \,
{{\mathcal{E}}_{j+1}}^{\delta}_{\beta} \nonumber
\end{eqnarray}
Writing down the 2-site Hamiltonian $\sum_j {\mathcal{H}}_{j \,
j+1}$ in a matrix representation $H_{j \, j+1}$, the relation with
the $\Rc$-matrix looks even simpler, namely
\begin{equation} (H_{2\,\mbox{\tiny sites}})^{\alpha
\beta}_{\gamma \delta}= \left.
\partial_u \check{R}^{\alpha \beta}_{\gamma \delta}(u) \right|_{u=u_0}
\label{2-site-Ham-mat2}
\end{equation}
\section{Polynomial R-matrix technique}
\label{secIII} In the previous section we showed that to each
solution $\Rc$ of the YBE (\ref{YBE}) one can associate a
fermionic Hamiltonian ${\mathcal{H}}={\mathcal{H}}_{j \, j+1}$
endowed with a set of symmetries ${\mathcal{J}}_n$'s. The
derivative of $\Rc$ with respect to $u$ basically represents the
two-site Hamiltonian ${\mathcal{H}}_{j \, j+1}$ through the
relations (\ref{2-site-Ham-op}) (operatorial form) or
(\ref{2-site-Ham-mat2}) (matrix representation).
\\The purpose is now to find solutions of the YBE, to figure out what
kind of Hamiltonian is associated to the solution and to determine
its physical features. Here below we describe a technique that we
recently developed to find solutions of the YBE, and then we
describe what kind of models we obtain.
\\Notice that if $\Rc(u)$ is a solution of (\ref{YBE}) then
$\Rc^{\prime}=f(u) \Rc(u)$, with $f(u)$ any scalar function
satisfying $f(0)=1$, is a solution as well, and it corresponds to
an Hamiltonian $\mathcal{H}^{\prime}=\mathcal{H}+c \, {\bf I}$
where $c=\frac{df}{d u} |_{u=0}$.\\

Our approach consists in looking for solutions of (\ref{YBE})
which are polynomials in the spectral parameter and satisfy
(\ref{cond-2}) for $u_0=0$. This implies:
\begin{equation}
\Rc(u)={\bf I}+u (H_{2\,\mbox{\tiny sites}}+c \, {\bf I})+
\frac{u^2}{2!} \Rc^{(2)}+ \ldots +\frac{u^p}{p !} \Rc^{(p)} \quad
, \label{Rc-ansatz}
\end{equation}
with $p$ the degree of the polynomial (unknown {\it a priori} and
possibly infinite). Here $\Rc^{(i)} \, (i=1, \ldots ,p)$ are
unknown coefficients which are to be fixed through eqn.(\ref{YBE})
and ${\bf I}$ the $d_V^2 \times d_{V}^2$ identity matrix.

Let us substitute the Ansatz (\ref{Rc-ansatz}) into eqn
(\ref{YBE}); due to the fact that YBE must be satisfied~$\forall
u, v$, we have to equate all the coefficients of the same power
$u^n \, v^m$. Thus we end up with {\it algebraic} equations, which
can be grouped order by order, the order being the value $l=m+n$.
The {\it Zero-}th and {\it First Order} equations are mere
identities which are always satisfied. For the {\it
\underline{Second Order}} we have
\begin{equation}
\Rc^{(2)}=(\Rc^{(1)})^2+\, {\bf I} \delta  \label{order2}
\end{equation}
where $\Rc^{(1)}=H+c \, {\bf I} \,$, and $\delta$, $c$ are
constants to be determined.\\

The {\it \underline{Third Order}} case still reduces to a single
equation, which reads
\begin{eqnarray}
\lefteqn{ \left\{ \Rc^{(1)}_{12} , \Rc^{(2)}_{23} \right\} -
\Rc^{(1)}_{23} \Rc^{(2)}_{23}-2
 \Rc^{(1)}_{23} \, \Rc^{(1)}_{12} \, \Rc^{(1)}_{23} +
\Rc^{(3)}_{23} =
} & & \nonumber \\
& & = \left\{ \Rc^{(1)}_{23} , \Rc^{(2)}_{12} \right\} -
\Rc^{(1)}_{12} \Rc^{(2)}_{12}-2
 \Rc^{(1)}_{12} \, \Rc^{(1)}_{23} \, \Rc^{(1)}_{12} +
\Rc^{(3)}_{12} \quad , \label{order3}
\end{eqnarray}
where $\{ \, , \, \}$ is the anticommutator.\\

The successive orders in general consist of more equations, up to
the {\it \underline{Highest Order}} $3 p$, in which case the
equation is simply the (spectral parameter independent) YBE,
\begin{equation}
\Rc^{(p)}_{23} \, \Rc^{(p)}_{12} \,  \Rc^{(p)}_{23} =
\Rc^{(p)}_{12} \, \Rc^{(p)}_{23} \,  \Rc^{(p)}_{12} \label{orderp}
\end{equation}

\noindent Notice that, as the second order equation (\ref{order2})
is explicitly solved once the Hamiltonian is given, the non
trivial equations start from the third order; for a polynomial of
degree $p \ge 1$ one has $3 p -2$ actual orders to consider.
\\The idea is to proceed by successive attempts: once the Hamiltonian
$\mathcal{H}$ is given, one first tries to solve the equations by
means of a first order polynomial $\Rc$~-matrix; if this is not
possible one passes to a second order polynomial and so on. If the
Ansatz (\ref{Rc-ansatz}) is correct one can hope to find the
$\Rc$-matrix in a finite number of steps, or guessing a recursive
formula for the coefficient $\Rc^{(i)}$. The advantage of this
method is that it deals with {\it algebraic} equations instead of
functional equations (cfr.(\ref{YBE})).

The simplest non-trivial case of a polynomial $R$-matrix is of
course obtained for $p=1$. In this case $\Rc(u)={\bf I}+u
\Rc^{(1)}$, and, due to (\ref{orderp}), $\Rc^{(1)}$ is nothing but
a braid operator. Moreover, eq. (\ref{order2}) implies that the
square of $\Rc^{(1)}$ is proportional to the identity (since
$\Rc^{(2)}=0$ for a first order polynomial)
\begin{equation}
(\Rc^{(1)})^2 \propto {\bf I} \label{squareid}
\end{equation}

Eqs.(\ref{orderp}) and (\ref{squareid}) imply that $\Rc^{(1)}$
(and hence the Hamiltonian $\mathcal{H}$) is in fact an element of
the symmetric group. The solutions of the symmetric group
relations have been fully investigated for a two-dimensional local
space (see for instance \cite{HIET}); whereas the case of
dimension $d_{V}=4$ of the local vector space $V$ was considered
in \cite{DOMO_SGE},\cite{DOMO_NUCLPHYSB}. In the next section we
extend the latter approach to generic dimension $d_{V}$.

\section{Generalized Permutator $R$-matrices}
\label{secIV} It is well known that, among the solutions of the
symmetric group relations (\ref{orderp})-(\ref{squareid}), one
always finds the permutation operator $P$, defined by $P
(e_\alpha\otimes e_\beta) = e_\beta \otimes e_\alpha$, where the
$e_\alpha$'s for $\alpha=1,\dots,d_{V}$ form an orthonormal basis
for the local vector space $V$. In fact, this is true whatever the
dimension of the representation is, and also holds for any {\it
graded} permutation operator $P_g$, where $P_g (e_\alpha\otimes
e_\beta)= (-)^{\epsilon(\alpha) \epsilon(\beta)} e_\beta\otimes
e_\alpha$, with $\epsilon_\alpha=0,1$ grading of the vector
$e_\alpha$. Therefore, both $P$ and $P_g$ give rise to integrable
models. One may wonder whether there are other solutions of
symmetric group relations generalizing the structure of $P$,
$P_g$. Here we propose the {\it generalized permutator} ones,
given by operators $\Pi$ which either transform one product of
basis vectors into the reversed product, or leave it unchanged,
according to a rule explained below. In the matrix form this means
that there is precisely one non-zero entry in each column and row
of $\Pi$. Moreover, due to (\ref{squareid}), the non-vanishing
entries of $\Pi$ must be equal to $+1$ or $-1$, up to an overall
multiplicative constant. Explicitly, $\forall \alpha,\beta =1,
\ldots d_{V} $, \be\label{genperm} \Pi (e_\alpha\otimes e_\beta) =
\underbrace{\theta^{d}_{\alpha\beta} (e_{\alpha} \otimes
e_{\beta})}_{\mbox{\tiny diagonal terms}} \, +
\,\underbrace{\theta^{o}_{\alpha\beta} \, (e_{\beta} \otimes
e_{\alpha})}_{\mbox{\tiny off-diagonal terms}} \quad .
 \ee Here
$\theta^{d}_{\alpha\beta}=\epsilon_{\alpha\beta}
s^{d}_{\alpha\beta}$, and
$\theta^{o}_{\alpha\beta}=(1-\epsilon_{\alpha\beta})
s^{o}_{\alpha\beta}$, where $\epsilon_{\alpha\beta}$ is a discrete
(0 or 1) valued function satisfying
$\epsilon_{\alpha\beta}=\epsilon_{\beta\alpha}$ and
$\epsilon_{\alpha\alpha}=1$, which selects the
diagonal/off-diagonal terms; moreover, $s^{d}_{\alpha\beta}=\pm 1$
accounts for additional signs of the diagonal entries, while
$s^{o}_{\alpha\beta}=\pm 1$ stands for the signs of off-diagonal
terms; we shall impose $\,s^{o}_{\alpha\beta}=s^{o}_{\beta\alpha}$
in order to make $\Pi$ a symmetric matrix.

It can be easily verified that the operators $\Pi$ have square
equal to the identity. Moreover, let us denote by ${\mathcal{S}}$
the set of index values, {\it i.e.}
${\mathcal{S}}=\{1,\dots,d_{V}\}$; the function
$\epsilon_{\alpha\beta}$ on ${\mathcal{S}} \times {\mathcal{S}}$
characterizes the structure of any $\Pi$. Let us denote by
${\mathcal{A}}^d$ the subset of $\mathcal{S} \times \mathcal{S}$
of the couples $(\alpha,\beta)$ for which
$\epsilon_{\alpha\beta}=1$: they determine the positions $4\,(
\alpha-1)+\beta$ of the non-vanishing {\it diagonal} entries;
since $\epsilon_{\alpha\beta}=\epsilon_{\beta\alpha}$, whenever
$\theta^{d}_{\alpha\beta} \neq 0$ we also have that
$\theta^{d}_{\beta\alpha} \neq 0$; in other words, the subset
${\mathcal{A}}^d$ can always be written in the form
${\mathcal{A}}^d= \bigcup_{i} \, {\mathcal{S}}_{i} \times
{\mathcal{S}}_{i}$ where the ${\mathcal{S}}_i$'s are disjoint
subsets of $\mathcal{S}$, with $i=1,\dots,N_S$, and $N_S\leq
d_{V}$ number of disjoint subsets in $S$ .
\\Inserting $\Pi$ into the relation (\ref{orderp}) it can be easily
found that it is a solution if and only if
$\theta^{d}_{\alpha\beta}=p_{i} \quad \forall \, (\alpha,\beta)
\in {\mathcal{S}}_{i} \times {\mathcal{S}}_{i}$, where $p_{i}=\pm
1$ are signs. The values of the $p_i$'s can be chosen
independently, and the remaining off-diagonal non-vanishing
entries are also free. In the following we shall refer to these
solutions as the generalized permutators, and with abuse of
notation we still denote them as $\Pi$.
\\Notice that this class of solutions exists for arbitrary $d_{V}$, and
in fact their actual number is an increasing function of $d_{V}$.
As an example, in the next sections we shall investigate how these
solutions provide $1440$ integrable models in case of fermionic
realizations with $d_{V}=4$ (unconstrained fermions with two
flavors), and $56$ models in case of two flavors constrained
fermions ($d_{V}=3$). At this stage, however, each of the
integrable models proposed has no free parameter; we shall discuss
in the next paragraphs how this limit can be partly removed by
adding to the generalized permutators conserved quantities with
free parameters. Of course, the actual number of the latter will
depend again on the dimension of the local vector space.

\section{Sutherland species and spectrum}
\label{secV} The meaning of a generalized permutator is easily
understood in terms of so called Sutherland species (SS). Starting
from a given local vector space $V$, we may think to group its
$d_V$ basis vector $e_{\alpha}$ (which in the sequel we shall
denote as physical species PS) into $N_S \leq d_V$ different
species ${\mathcal{S}}_i$ which we call Sutherland species. Each
of these species is left unchanged by the action of the
generalized permutator, the latter interchanging only basis vector
belonging to different SS's. A generalized permutator would then
have the structure of an ordinary permutator if represented on a
local vector space of dimension $N_S$. Notice that, with respect
to this interpretative scheme, apparently the solutions presented
in the previous section allow some extra freedom in the choice of
signs of non-diagonal elements; on the contrary, here the
off-diagonal elements connecting different PSs of the same SS
$S_i$ to a different SS $S_j$ must have the same sign, say $s_i$.
Such difference disappears through a redefinition of phase of some
basis vectors belonging to $S_i\otimes S_j$.

The only signs in (\ref{genperm}) which cannot be changed by a
mere redefinition of basis are those of diagonal elements. Indeed,
these signs can be used to classify the different $\Pi$ solutions
of GYBE for a given dimension $d_V$. Following Sutherland's
notation\cite{SUT}, each species can be classified as either
`fermionic'($F$) or `bosonic'($B$), according to the sign of
diagonal elements. Hence all integrable models given in
(\ref{genperm}) are of type \be F^{N_S-l}B^l \quad , \quad
l=0,\dots,N_S \quad , \quad 2\leq N_S\leq d_V \quad ,\ee and
characterize a model with $N_S-l$ fermionic and $l$ bosonic SS's.
Different models that are recognized to be of the same SS-types
also share the same structure of Coordinate or Algebraic Bethe
Ansatz equations. The number of different types of such structures
associated with the generalized permutators can be easily
evaluated, and is \be\nu (d_V) ={1\over 2} d_V^2+{3\over 2} d_V -2
\quad ,\ee  for given $d_V$. We remark that this is not to be
confused with the number of different integrable Hamiltonians
which in general is greater.

The cases $F^{P}$ and $B F^{P}$ have have been investigated in
\cite{SUT} within the framework of the Coordinate Bethe Ansatz,
and ground state energy equations have been obtained. Moreover,
the cases $F^P$ and $B^P$ have been explicitly examined in
\cite{KURE} within the QISM; and some other cases ($B F^2$ and
$B^2 F^2$) share the same algebraic structure as known models
($t-J$\cite{FOKA,EKStJ} and $EKS$\cite{EKS} respectively).
\\In particular, for the $BF$ type of models, the entire spectrum
can be obtained, and coincide, up to constant terms, with that of
spinless fermions on an open chain:
\begin{equation}
E(\{n_l \})=-2 \sum_{l=1}^{L} \cos \left(\frac{\pi l}{L+1}\right)
\, n_l \quad , \label{spectrum}
\end{equation}
where $n_l=0,1$ are quantum numbers and $L$ the length of the
chain.
\\Notice that in the latter case we have the spectrum for any
$d_V$. This means that in particular the ground state energy is
independent of $d_V$. However, it must be realized that the actual
degeneracy of each eigenvalue depends on the way the Sutherland's
species are realized in terms of physical species for each
specific model. This enters explicitly the calculation of both the
ground state energy and the partition function, determining
different physical features of models with the same $FB$
structure, as we shall discuss for the case $d_V=4$ in a
subsequent section.
\\Within the spectrum, the ground state is particularly worth of interest,
because these Hamiltonians are expected to well describe materials
that exhibit peculiar physics at low temperatures. For the $F^P$
case, it can be shown \cite{SUT} that the miminum of the energy,
whose eigenvalue reads
\begin{equation}
\epsilon_0=1-\frac{2}{P} \int_{0}^{1}   dx \,
\frac{x^{\frac{1}{P}-1}-1}{1-x} \quad \quad , \label{eps-FP}
\end{equation}
is reached at equal densities of all fermionic species.
\\Also, again as far as the ground state is concerned, it
has been shown \cite{DOMO_RC} that Sutherland's theorem
(originally formulated for permutators of physical species) can be
extended to the generalized permutators, and it is thus possible
to assert that, in the thermodynamic limit, the ground state
energy (per site) $\epsilon_0$ of a $B^n F^m$ problem is equal to
that of a $B F^m$ problem, for which the Bethe Ansatz equation
have been formulated in full generality in \cite{SUT}.

\section{Applications to extended Hubbard Models}
\label{secVI} Here we review how the generalized permutator
approach succeeds in producing a class of integrable models when
applied to Hubbard-like systems of interacting electrons.
\subsection{Hamiltonian and representation}
We consider here a quite general 1-band extended isotropic Hubbard
model preserving the total spin and number $N$ of electrons, which
reads
\end{multicols}
\begin{eqnarray}
&{\cal H}& = - \sum_{\langle{j},{k}\rangle,\sigma }[t-X
(\hat{n}_{{j}, -\sigma} + \hat{n}_{{k},-\sigma})+\tilde X
\hat{n}_{{j},-\sigma} \hat{n}_{{k},-\sigma} ]
c_{{j},\sigma}^\dagger c_{{k}, \sigma} +  U \sum_{j}
\hat{n}_{{j},\uparrow}\hat{n}_{{j}, \downarrow} + {V\over
2}\sum_{\langle{j},{k}\rangle} \hat{n}_{j}
\hat{n}_{k} -\mu \sum_{j} \hat{n}_j \label{EHM}  \\
&+& {W\over 2} \sum_{\langle{j},{k}\rangle,\sigma,\sigma' }
c_{{j},\sigma}^\dagger c_{{k}, \sigma'}^\dagger c_{{j},{\sigma}'}
c_{{k}, \sigma} + Y\sum_{\langle{j},{k}\rangle }
c_{{j},\uparrow}^\dagger c_{{j},\downarrow}^\dagger
c_{{k},\downarrow} c_{{k}, \uparrow} + P
\sum_{\langle{j},{k}\rangle } \hat{n}_{{j},\uparrow} \hat{n}_{{j},
\downarrow} \hat{n}_{k} + {Q\over 2} \sum_{\langle{j},{k}\rangle }
\hat{n}_{{j},\uparrow} \hat{n}_{{j}, \downarrow}
\hat{n}_{{k},\uparrow} \hat{n}_{{k}, \downarrow} \, , \nonumber
\end{eqnarray}
\begin{multicols}{2}
\noindent In (\ref{EHM}) $t$ represents the hopping energy of the
electrons, while the subsequent terms describe their Coulomb
interaction energy in a narrow band approximation: $U$
parametrizes the on-site repulsion, $V$ the neighboring site
charge interaction, $X$ the bond-charge interaction, $W$ the
exchange term, and $Y$ the pair-hopping term. Moreover, additional
many-body coupling terms have been included in agreement with
\cite{dBKS}: $\tilde X$ correlates hopping with on-site occupation
number, and $P$ and $Q$ describe three- and four-electron
interactions. Finally, $\mu$ is the chemical potential. \\The
local vector space $V$ at each lattice site has $d_V=4$, and in
the following we shall identify the 4 physical states $|\uparrow
\rangle$, $|\downarrow \rangle$, $| 0 \rangle$ and $|\downarrow
\uparrow \rangle$ with the canonical basis $e_{\alpha}$ of ${C}^{
4}$.

The matrix representation of the two-site Hamiltonian ${\cal
H}_{j,j+1} $ can be obtained through the projectors
(\ref{Hubb-proj}) in the following way
\begin{equation}
{\mathcal{H}}_{j \, j+1}=(-1)^{p(\gamma) (p(\beta)+p(\delta))}
 (H_{EH}^{(2)} )^{\alpha \beta}_{\gamma \delta} \, \,
{{\mathcal{E}}_j}^{\gamma}_{\alpha} \,
{{\mathcal{E}}_{j+1}}^{\delta}_{\beta} \label{2-site-Ham-op}
\end{equation}
Explicitly
\end{multicols}
\begin{equation}
H_{EH}^{(2)}  = {\scriptsize{\pmatrix{
\:h^{11}_{11} & 0 & 0 & 0 &|& 0 & 0 & 0 & 0 &|& 0 & 0 & 0 & 0 &|&
0 & 0 & 0 & 0 &\cr
\:0 & h^{12}_{12} & 0 & 0 &|& h^{12}_{21} & 0 & 0 & 0 &|& 0 & 0 &
0 & h^{12}_{34} &|& 0 & 0 & h^{12}_{43} & 0 &\cr
\:0 & 0 & h^{13}_{13} & 0 &|& 0 & 0 & 0 & 0 &|& h^{13}_{31} & 0 &
0 & 0 &|& 0 & 0 & 0 & 0 &\cr
\:0 & 0 & 0 & h^{14}_{14} &|& 0 & 0 & 0 & 0 &|& 0 & 0 & 0 & 0 &|&
h^{14}_{41} & 0 & 0 & 0 \cr
- & - & - & - &|& - & - & - & - &|& - & - & - & - &|& - & - & - &
- &\cr
0 & h^{12}_{21} & 0 & 0 &|& h^{21}_{21} & 0 & 0 & 0 &|& 0 & 0 & 0
& h^{21}_{34} &|& 0 & 0 & h^{21}_{43} & 0 &\cr
0 & 0 & 0 & 0 &|& 0 & h^{22}_{22} & 0 & 0 &|& 0 & 0 & 0 & 0 &|& 0
& 0 & 0 & 0 &\cr
0 & 0 & 0 & 0 &|& 0 & 0 & h^{23}_{23} & 0 &|& 0 & h^{23}_{32} & 0
& 0 &|& 0 & 0 & 0 & 0 &\cr
0 & 0 & 0 & 0 &|& 0 & 0 & 0 & h^{24}_{24} &|& 0 & 0 & 0 & 0 &|& 0
& h^{24}_{42} & 0 & 0 \cr
- & - & - & - &|& - & - & - & - &|& - & - & - & - &|& - & - & - &
- &\cr
0 & 0 & h^{13}_{31} & 0 &|& 0 & 0 & 0 & 0 &|& h^{31}_{31} & 0 & 0
& 0 &|& 0 & 0 & 0 & 0 &\cr
0 & 0 & 0 & 0 &|& 0 & 0 & h^{23}_{32} & 0 &|& 0 & h^{32}_{32} & 0
& 0 &|& 0 & 0 & 0 & 0 &\cr\
0 & 0 & 0 & 0 &|& 0 & 0 & 0 & 0 &|& 0 & 0 & 0 & 0 &|& 0 & 0 & 0 &
0 &\cr
0 & h^{12}_{34} & 0 & 0 &|& h^{21}_{34} & 0 & 0 & 0 &|& 0 & 0 & 0
& h^{34}_{34} &|& 0 & 0 & h^{34}_{43} & 0 \cr
- & - & - & - &|& - & - & - & - &|& - & - & - & - &|& - & - & - &
- &\cr
0 & 0 & 0 & h^{14}_{41} &|& 0 & 0 & 0 & 0 &|& 0 & 0 & 0 & 0 &|&
h^{41}_{41} & 0 & 0 & 0 &\cr
0 & 0 & 0 & 0 &|& 0 & 0 & 0 & h^{24}_{42} &|& 0 & 0 & 0 & 0 &|& 0
& h^{42}_{42} & 0 & 0 &\cr
0 & h^{12}_{43} & 0 & 0 &|& h^{21}_{43} & 0 & 0 & 0 &|& 0 & 0 & 0
& h^{34}_{43} &|& 0 & 0 & h^{43}_{43} & 0 &\cr
0 & 0 & 0 & 0 &|& 0 & 0 & 0 & 0 &|& 0 & 0 & 0 & 0 &|& 0 & 0 & 0 &
h^{44}_{44} }}} \label{matrix}
\end{equation}
where
\begin{eqnarray}
h^{11}_{11}&=&h^{22}_{22}=\mu+V-W  \hspace{1.9cm}
h^{12}_{12}=h^{21}_{21}=\mu+V \hspace{1.1cm}
h^{13}_{13}=h^{31}_{31}=h^{23}_{23}=h^{32}_{32}= \mu /2
\nonumber  \\[0.3cm]
h^{12}_{21}&=&W \hspace{1.5cm} h^{34}_{43}=Y  \hspace{1.5cm}
h^{13}_{31}=h^{23}_{32}=-t \hspace{1.6cm}
h^{14}_{41}=h^{24}_{42}=t-2 X+ \tilde{X}
\nonumber \\[0.3cm]
h^{14}_{14}&=&h^{41}_{41}=h^{24}_{24}=h^{42}_{42}=\frac{3}{2} \,
\mu +P+\frac{U}{2}+2 V-W  \hspace{1.9cm}
h^{34}_{34}=h^{43}_{43}=\mu+\frac{U}{2}  \label{matrix-entries}
\\[0.3cm]
h^{44}_{44}&=&2 \mu+ 4 P+Q+U+4 V-2 W \hspace{1.5cm}
h^{12}_{34}=h^{12}_{43}=-h^{21}_{34}=-h^{21}_{43}=t-X \quad .
\nonumber
\end{eqnarray}

\subsection{Integrable cases}
We would like to recognize in the matrix (\ref{matrix}) one of the
generalized permutators given in section $3$. According to the
scheme discussed there, this would guarantee that the
corresponding Hamiltonian is then integrable. We first observe
that this correspondence is possible only if $t-X=0$; indeed only
in this case the non vanishing entries of $H_{EH}^{(2)}$
 may coincide
with those of a generalized permutator. When this condition is
implemented, the representation of the two-site extended Hubbard
Hamiltonian (up to an additive constant $c$) is itself a
generalized permutator whenever some linear relations among the
non-vanishing Hamiltonian entries and the non-vanishing elements
of a generalized permutator are satisfied. It turns out that
actually there are $96$ different possible choices of values of
the physical parameters in ${\cal H}$ satisfying such relations.
They can be cast into six groups as (\ref{hsol}) shows.
\begin{equation}
\scriptsize{\matrix{
  &  &  H_1 (s_1,\dots, s_5) & | & H_2 (s_1,\dots,s_5) &|&
        H_3 (s_1,s_2,s_3)&| & H_4 (s_1,s_2,s_3) &|&
        H_5 (s_1,s_2,s_3)&| & H_6 (s_1,s_2,s_3) &|& \cr
  &||& ------& | &------& | & ------ & | & ------& | & ------ & | & ------
& |\cr t &||&  1      &|&  1  &| &  1 &|      & 1 &|& 1  &|& 0 &|
\cr X &||&  1      &|&  1  &| &  1 &|      & 1 &|& 1  &|& 0 &| \cr
\tilde{X} &||&  1+s_2 &|& 1+s_2 & |& 1+s_2 &|& 1+s_2 &|& 1 &|& 1
&|\cr U &||& 2 s_1&| & 2 s_1 &|& 4 s_1 &|& 4 s_1 &|& 2 s_1 &|& -2
s_1 &|  \cr V &||& s_1 &|& s_1+s_4 &|& s_1   &|& s_1+s_3 &|&
s_1+s_3  &|& 0 &|\cr W &||& s_4 &|& 0 &|& s_3 &|& 0 &|& 0  &|& 0
&|\cr Y &||& s_3 &|& s_3  &|  & 0 &|& 0 &|& s_2  &|& s_2 &|\cr P
&||& s_4-s_1 &|&  -s_1-2 s_4 &|& s_3-2 s_1 &|& -2 (s_1+s_3) &|&
-(s_1+s_3) &|&  0  &| \cr Q &||&  -2 s_4+s_1+s_5 &|& 4s_4+s_1+s_5
&|& 4 s_1-2 s_3 &|& 4 (s_1+s_3)
 &|&  s_1+s_3   &|&  s_1+s_3  &| \cr
\mu &||& -2 s_1&| & -2 s_1&| & -2 s_1&| & -2 s_1 &|& -2 s_1  &|& 0
&| \cr }} \label{hsol}
\end{equation}
\begin{multicols}{2}
Here $s_i=\pm 1$, $i=1,\dots, 5$ are arbitrary signs; the first
two groups consist of $32$ different solutions each, while any of
the other four groups is made of $8$ different cases. Notice that
requiring the conservation of spin has reduced the $1440$
generalized permutators solutions of YBE for $d_V=4$ down to the
present $96$ solutions described in (\ref{hsol}). \\As expected,
$t=X$ is a common feature exhibited by all the solutions, implying
that the number of doubly occupied sites is a conserved quantity
for those ${\cal H}$ that are derivable from first-degree
polynomial $R$-matrices. This feature is important in that it
means that in these cases ${\cal H}$ can be diagonalized within a
sector with a given number of up and down electrons and doubly
occupied sites. In practice, the solvability of the model in one
dimension is not affected by having values of $U$ and $\mu$ in
${\cal H}$ other than those reported in (\ref{hsol}) (see also
\cite{EKS}), as well as by the presence of an external magnetic
field ({\it i.e.} adding to ${\cal H}$ a term proportional to
$\sum_j n_{j,\uparrow}-n_{j,\downarrow}$).

The $96$ integrable cases given in (\ref{hsol}) can be classified
according to the scheme of SS's, which allow to distinguish just
12 different algebraic structures. In particular, the first group
is characterized by $N_S\equiv d_V (=4)$ SS's, hence all models
are of type $F^{4-l} B^l$; whereas the second and the third group
correspond to $N_S=3$ SS's, and the remaining three groups all
correspond to $N_S=2$ SS's. To all these structures are associated
different sets of Bethe Ansatz equations, which have been
explicitly worked out in \cite{AUST}, even though not explicitly
solved. In what follows we illustrate a different approach to some
of these algebraic structures, which allow an explicit evaluation
of many physical features.

\subsection{Ground state and thermodynamics}
\begin{figure}
\epsfig{file=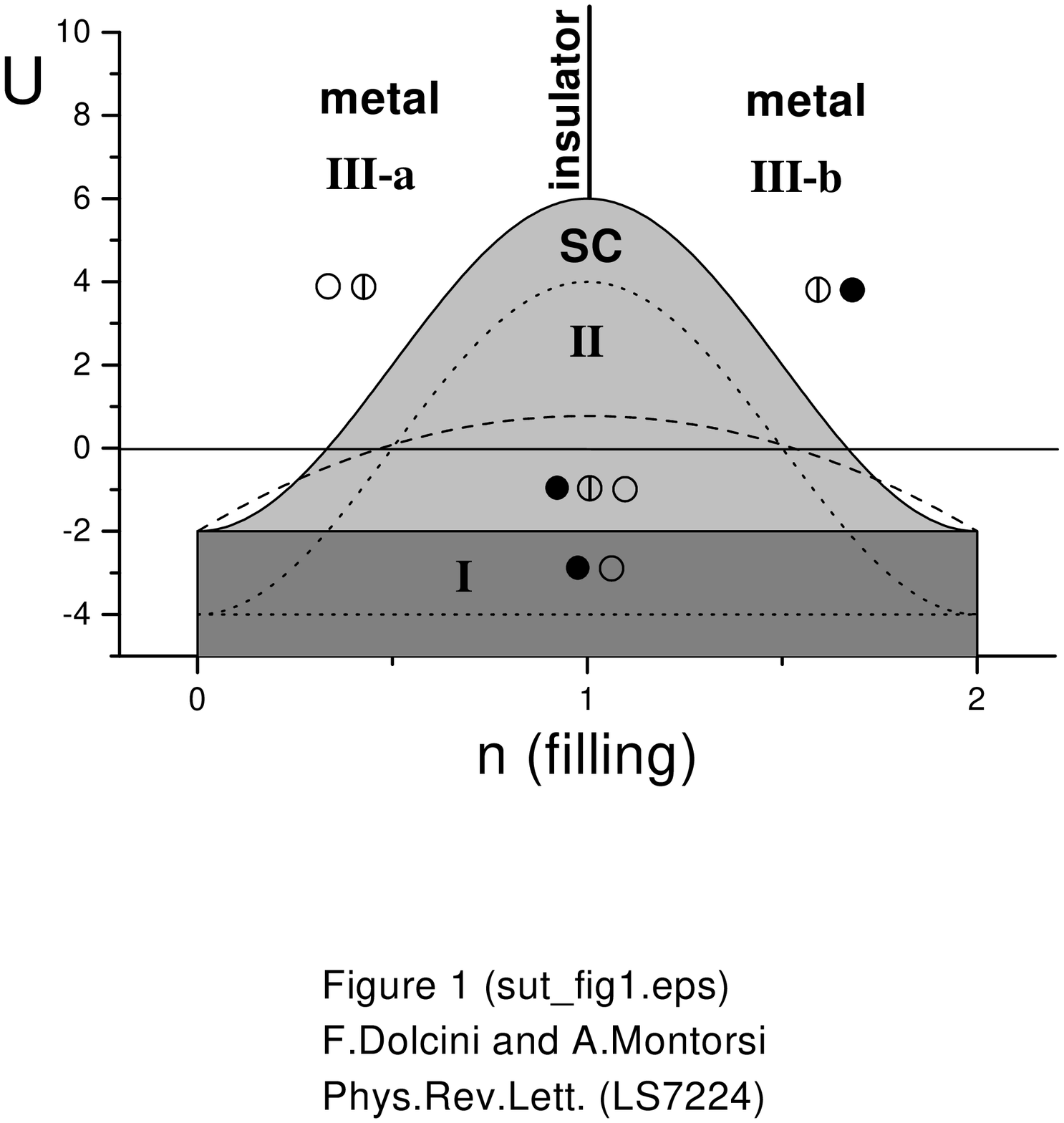,scale=0.4,clip=} \caption{ Ground state
phase diagram of models with a single doubly degenerate SS of type
F; for all models $t=X=1$. Moreover the continuous lines describe
the case $\tilde{X}=(1-\sigma)$; $Y=-\sigma$; $P=-1$; $Q=2$ [28] ;
the dashed line is the EKS model [27] ($\tilde{X}=W=Y=V=1$,
$P=Q=0$), and the dotted line corresponds to the AAS model
($\tilde{X}=W=Y=V=P=Q=0$ [31] ). All models exhibit an
insulator-superconductor transition at $n=1$, for different
$U_c$.\label{sut_fig1}}
\end{figure}

\begin{figure}
\epsfig{file=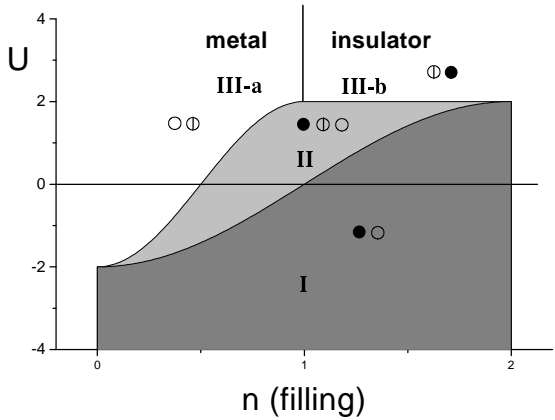,scale=0.4,clip=} \caption{Ground state
phase diagram of the model $X=\tilde{X}=1$ ; $Y=\sigma$;
$V=W=P=Q=0$. A filling controlled metal-insulator transition takes
place for $U \ge 2$  between two {\it finite} regions, III-a and
III-b.\label{sut_fig2}}
\end{figure}

Recognizing that a model identifies (up to some commuting terms) a
set of Sutherland Species greatly simplifies the calculation of
the spectrum. The crucial point which allows that is the use of
{\it open} boundary conditions, instead of the customary periodic
ones; although in the thermodynamic limit the bulk properties are
not affected by either choice, the calculations are more
straightforward for the former. Indeed in an open one-dimensional
chain the set of eigenvalues of a generalized permutator is equal
to that of an ordinary permutator between $N_S$ objects, i.e. the
effective dimensionality of the Hilbert space is reduced
(reduction theorem). As a consequence of that, the degeneracy of
the eigenvalues can also be computed, simply counting the ways one
can realize a given configuration of Sutherland Species.
\\The reduction theorem is proved when realizing that, according
to what observed above, the relative order of any sequence of
states belonging to the same species is preserved\cite{AAS}; the
Hamiltonian can therefore be diagonalized within each subspace of
given set of sequences. For instance, for $N_S=2$ any model of FB
type has a spectrum given by (\ref{eps-FP}) (plus conserved
quantities), and the ground state energy per site is simply
obtained by filling all the lower energy levels up to the Fermi
level; it turns out to be just a function of the total number of
$F$ particles in the ground state, $n_F$, and it reads \be
\epsilon_0 (n_F) = 2 n_F-{2\over \pi} sin(\pi n_F)+(U-\bar
U)n_{\uparrow,\downarrow} +(\mu-\bar\mu) n \quad ; \label{gse}\ee
here $\bar U$, and $\bar \mu$ are the value of $U$ and $\mu$ as
fixed from table (\ref{hsol}), $n={1\over L}\sum_j \hat n_j$, and
$n_{\uparrow,\downarrow}\doteq {1\over L}\sum_{j}
\hat{n}_{{j},\uparrow}\hat{n}_{{j}, \downarrow}$ is itself a
function of $n_F$, the explicit form of which depends on the
specific FB model we are looking at. Hence, the true ground state
energy of the model is obtained by minimizing (\ref{gse}) with
respect to $n_F$, and will depend on the physical parameter $n$
and $U$. This is shown in fig (\ref{sut_fig1})-(\ref{sut_fig2}).
There we plot the ground state phase diagram for different FB
models, in which $t=X=1$, $U$ and $n$ are varied, and the other
interaction parameters are chosen in different specified ways, so
as to have, within the same $F$ species, two PS  for
(\ref{sut_fig1}), and three PS for (\ref{sut_fig2}). \\The empty
circles represent empty states, the barred circles represent
singly occupied states, the full circles are doubly occupied
states; regions in which there are no singly occupied states are
characterized by $n_F=0$.
The dramatic differences in the phase diagram of models
represented in the figures are simply due to the fact that in the
first case  the F species contains two physical species
($|\uparrow\rangle$, and $|\downarrow\rangle$), implying
$n_{\uparrow,\downarrow}={1\over 2} (n_F-n)$, and the B species
contains the other two ($|\uparrow\downarrow\rangle$, and
$|0\rangle$); whereas in the second case species F consists of one
more physical species ($|\uparrow\downarrow\rangle$, hence
$n_{\uparrow,\downarrow}=n-n_F$), and B is just the empty state.
Such difference also reflects on the degeneracy of each level of
the spectrum, and ultimately manifests in a different behavior of
thermodynamical quantities.
\\For instance, let us derive explicitly the partition function
for one of the models in fig. 1, corresponding to the Hubbard
model with fixed bond-charge interaction ($X=t$) (also known as
AAS model \cite{AAS}). Implementing the relation between $n_F$ and
$n_{\uparrow\downarrow}$, the spectrum (\ref{spectrum}) in
presence of Coulomb repulsion, chemical potential, and external
magnetic field becomes
\begin{eqnarray}
E&=&E(\{n^{(F)}_{k}\};N_\uparrow;N_{\uparrow\downarrow})= \label{SPE} \\
&=&\sum_{k} (\epsilon_k-\mu+h) n_k^{F} \,  + (U-2\mu)
N_{\downarrow \uparrow}-2h N_{\uparrow} \nonumber
\end{eqnarray}
where $\epsilon_k=-2t \cos k$.
\\The degeneracy $g$ corresponds to the different ways one
can realize a configuration of Sutherland species, with the
constraint that the total numbers $N_{\downarrow \uparrow}$ and
$N_\uparrow$ appearing in (\ref{SPE}) remain unchanged; a simple
calculation yields
\begin{equation}\label{deg}
g(E(\{N_{F}\};N_{\downarrow \uparrow};N_{\uparrow}))={L-N_F
\choose N_{\downarrow \uparrow}} \, {N_F \choose N_{\uparrow}}
\end{equation}
The rearrangement of the Fock space deriving from the
identification of the Sutherland Species allows a straightforward
calculation of the (gran-canonical) partition function
\begin{eqnarray}
&{\mathcal Z} & = \sum_{ \{ n^{{F}}_k \} } \, \sum_{N_{\downarrow
\uparrow}=0}^{L-N_F} \sum_{N_\uparrow=0}^{N_F} \, g(E) e^{-\beta E
(\{n^{{F}}_{k}\};N_{\downarrow \uparrow};N_{\uparrow}) }\,
 = \label{zeta} \\
 & = & (1+e^{\beta(\mu-\frac{U}{2})})^{L} \prod_{k=1}^L
\left( 1+ e^{\left[- \beta \, (\epsilon_k-\mu^{*}(\mu,\beta,U,h))
\, \right]}  \right) \nonumber
\end{eqnarray} \noindent
In the second line of (\ref{zeta}) we have defined
\begin{equation}
\mu^{*}(\mu,\beta,U,h)= \mu +\frac{1}{\beta} \ln \frac{2 \cosh
\beta h}{1+\exp{2 \beta (\mu -U/2)}} \label{mus}
\end{equation}
$\beta=1/(k_B T)$ being the inverse temperature. Notice also that
the product over $k$ resulting in (\ref{zeta}) is in form similar
to the partition function of a tight binding model of spinless
fermions, where $\mu^{*}$ plays the role of an effective chemical
potential renormalized by the interaction $U$, the magnetic field
$h$ and the temperature itself.

By means of the partition function (\ref{zeta}), one can calculate
the thermodynamic observables from the gran potential (per site)
$\omega=-\lim_{L\rightarrow \infty} k_B T \ln{\mathcal{Z}}$. This
is done in \cite{DOMO_PRB}. In the following, we review the
results concerning the specific heat.

\begin{figure}
\epsfig{file=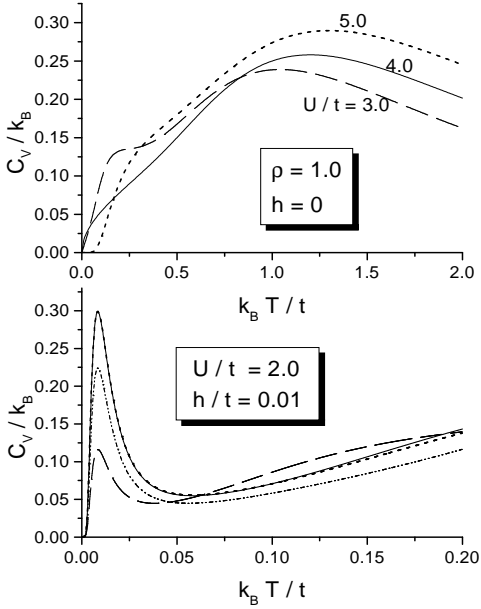,width=6.5cm,height=8.5cm,clip=}
\caption{The specific heat as a function of temperature. Top:
$\rho=1$, $h=0$: the metal-insulator transition is revealed
through the change in low-temperature behavior from linear to
exponential across the critical value $U=4t$. \\Bottom: $C_V$ for
different filling values: $\rho=0.25$ (dashed), $\rho=0.50$
(dot-dashed); $\rho=0.75$ (dotted) and $\rho=1$ (solid); a
low-temperature sharp peak emerges for non-vanishing magnetic
field.} \label{aas_fig1}
\end{figure}

In fig.\ref{aas_fig1} we have plotted the specific heat (per site)
$C_V$ as a function of the temperature. Here the chemical
potential $\mu$ is eliminated in favor of the filling through the
relation $\rho=(n=)\partial\omega/\partial \mu$, as usual.
 In particular, in the top figure we have examined the case of
 half filling (i.e. $\rho=1$) and zero magnetic field ($h=0$), for different values
of the on-site Coulomb repulsion $U$. One can observe that, across
the value $U/t=4$, the low-temperature behavior of $C_V$ changes
from linear to exponential; explicitly, for $U<4t$ we have
\begin{equation}
C_V \sim \gamma T \hspace{0.8cm} \mbox{with} \hspace{0.8cm}
\gamma= \frac{ k_B^2 \pi}{6 t \, \sqrt{(1-(U/4t)^2)}} \, \, ,
\label{gamma1}
\end{equation}
whereas for $U>4t$
\begin{equation}
C_V \, \sim k_B \, \frac{(U-4t)^2}{8 \sqrt{\pi} t^2} (k_B
T/t)^{-3/2} \, e^{-\frac{(U-4t)}{2K_B T}} \quad. \label{exp1}
\end{equation}
This is a finite-temperature effect of a metal-insulator
transition, in accordance with the result obtained in \cite{AAS},
where a {\it charge} gap $\Delta_c=U-4t$ is shown to open in the
ground state for $U>4t$. We recall that for $X=0$ (i.e. for the
ordinary 1D Hubbard model) no metal-insulator transition occurs;
the bond charge term thus seems to give rise to a finite critical
value $U_c$, increasing from 0 to $4t$ as the coupling $X$ is
varied from 0 to t.
In the bottom fig.\ref{aas_fig1}, $C_V$ is plotted for different
filling values, fixed ratio $U/t=2$ and magnetic field $h/t=0.01$.
A sharp low-temperature peak, located at $k_B T \sim h$, is
observed to emerge as soon as the magnetic field is turned on.
\\One can show that
\begin{equation}
\lim_{T\rightarrow 0} \lim_{h\rightarrow 0} C_V/T \neq
\lim_{h\rightarrow 0} \lim_{T\rightarrow 0} C_V/T \label{non-int}
\end{equation}
differently from the ordinary Hubbard model, where the two limits
are interchangeable\cite{TAKA}. At half-filling and for
$|U+|2h||<4t$, for instance, one has $C_V \sim \gamma \, T $ with
\begin{equation}
\gamma= \frac{ k_B^2 (3 \ln^2 2+\pi^2)}{6 \pi t \,
\sqrt{(1-((U+2|h|)/4t)^2)}} \label{gamma2}
\end{equation}
Comparing eq.(\ref{gamma2}) to eq.(\ref{gamma1}), one can realize
that (\ref{non-int}) holds. This can be interpreted as a signal of
non Fermi liquid behavior. Similarly, the exponential behavior,
occurring when the gap is open, is different; namely, for
$|U+|2h||>4t$
\begin{equation}
C_V \sim k_B \frac{(U+2|h|-4t)^2}{4 \sqrt{\pi} t^2} (k_B
T/t)^{-3/2} \, e^{-\frac{(U+2|h|-4t)}{2K_B T}} \quad.
\end{equation}
to be compared to eq.(\ref{exp1}).

\begin{figure}
\epsfig{file=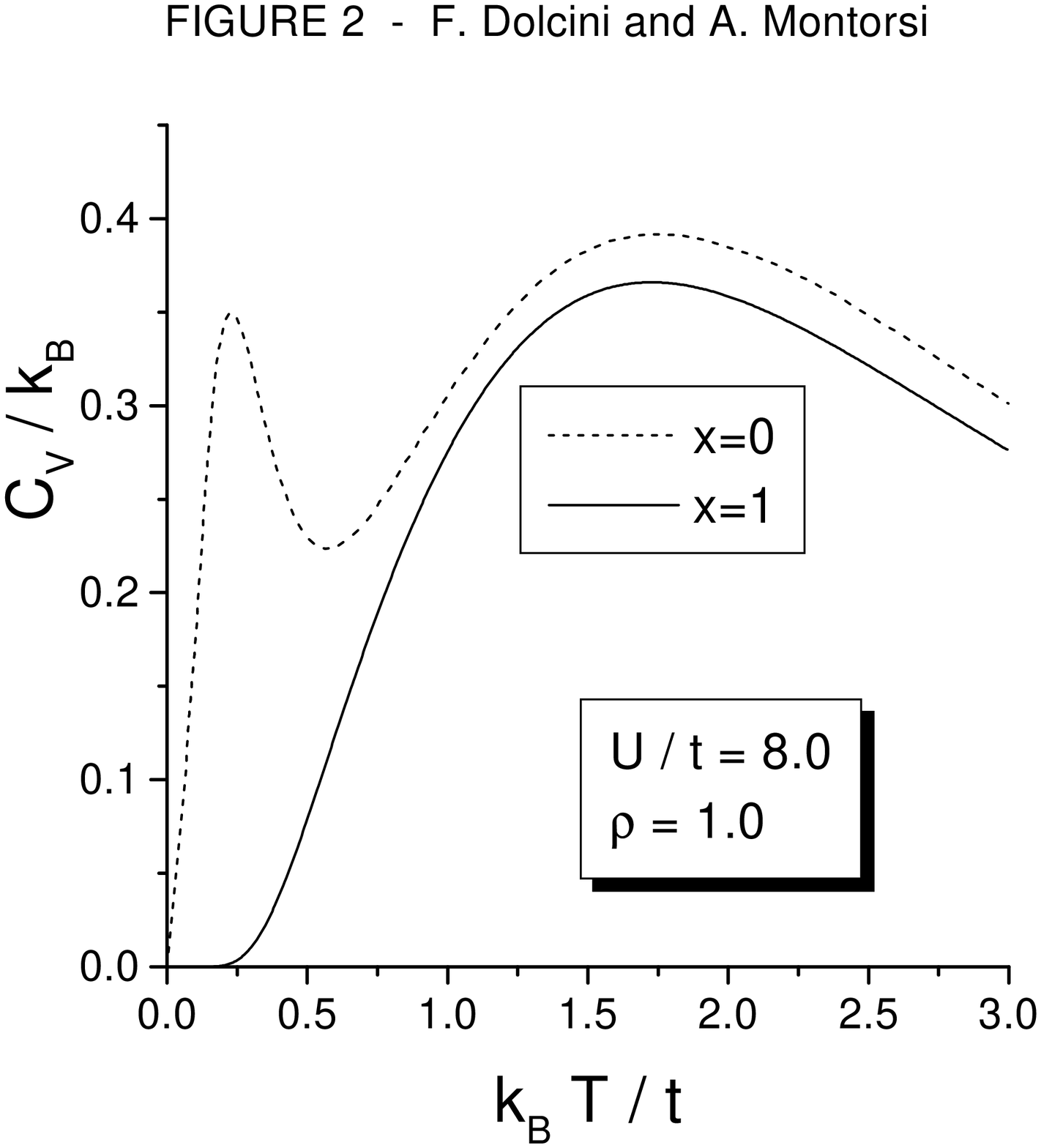,width=6.5cm,height=6.5cm,clip=}
\caption{The specific heat as a function of temperature for model
the bond-charge model in the strong coupling regime ($U=8t$), at
half filling and zero magnetic field. The dotted line is the case
$X=0$ - i.e. the ordinary Hubbard model- obtained from [16], and
the solid line the case $X=t$, obtained from our exact
calculations. Continuity arguments suggest that the specific heat
for arbitrary $0 \le X \le t$ lies between these two curves. The
low-temperature peak originating from spin excitations is depleted
by the bond-charge interaction.} \label{aas_fig2}
\end{figure}

In fig.\ref{aas_fig2} the specific heat of the Hubbard model with
bond charge is plotted for $X=1$ is aside the case $X=0$ (i.e. the
Hubbard model) for strong coupling, namely $U=8t$.
Notice that the ordinary Hubbard model has a low-temperature peak,
whose origin is due to spin degrees of freedom; the latter being
not gapped, the low-temperature behavior of $C_V$ is linear in
spite of the fact that a charge gap is present at any
$U>0$\cite{LIWU}. In contrast, for $X=t$ the spectrum does not
carry any spin quantum number, due to the rich symmetry of the
model; spins act therefore as a sort of dummy variables. Although
the value $X=t$ is a particular one, it is reasonable to expect
that, for continuity argument, the plot of $C_V$ for intermediate
values $0 \le X \le 1$ lies between the two curves. As a
consequence, we can infer that the effect of spin excitations is
weakened by the presence of the bond-charge interaction, at least
in the strong coupling regime.

\section{Applications to constrained fermions models}
\label{secVII} In this section we would like to discuss a simpler
application of the above scheme, which is the case of two flavors
electrons (up and down spin, for instance) with the constraint
that no two electrons can occupy the same site (constrained
fermions). The local vector space in this case has dimension
$d_V=3$, the physical species at each site being just
$|\uparrow\rangle$, $|\downarrow\rangle$, and $|0\rangle$.
\\The more general constrained fermions Hamiltonian with nearest
neighbor interaction one may think of is an extended $t-J$
Hamiltonian\cite{tJ}, define as \bq H_{EtJ}&=& - t
\sum_{j,\sigma}(\tilde c_{j,\sigma}^\dagger \tilde
c_{j+1,\sigma}+h.c.) + J\sum_j \vec{{\mathcal{S}}}_{j} \cdot
\vec{{\mathcal{S}}}_{j+1}\cr &+& V \sum_j \tilde n_j \tilde
n_{j+1} \quad ,\label{etj} \eq where $\tilde c_{j\sigma}\doteq
(1-\hat n_{j,\bar\sigma}) c_{j,\sigma}$, $\tilde n_{j\sigma}\doteq
(1-\hat n_{j,\bar\sigma})\hat n_{j\sigma}$, $\bar\sigma=-\sigma$,
and terms not conserving the total number of electrons and the
total spin operator have been ignored. Interestingly, $H_{EtJ}$
reduces to the infinite $U$ Hubbard model for $J=V=0$, and to the
standard $t-J$ model for $V=-{1\over 4}$, both of which have been
solved in one dimension.
\\Again, in order to answer the question whether there are
$H_{EtJ}$ Hamiltonians of generalized permutator type, we have to
represent $H_{EtJ}$ as a matrix, and this can be done by using the
Hubbard projectors introduced in section 2 (the constraint of no
double occupancy identifies in fact the $3\times 3$ upper left
submatrix of (\ref{Hubb-proj})). Interestingly, it turns out that
the representation of the $2$ sites Hamiltonian identified by
(\ref{etj}) in this case has non-vanishing off-diagonal entries
precisely where those of generalized permutators are. However, by
solving the linear equations in the interaction parameters
steaming from identifying such non-vanishing entries, it turns out
that there are just $8$ different integrable
${\mathcal{H}}_{EtJ}$, to be compared with the $56$ different
generalized permutator obtained for $d_V=3$; the reduction of
solutions being due to the required conservation of spin and
charge operators.

The integrable cases can be classified according to the total
number of corresponding SS into the following two classes:
\end{multicols}
\begin{eqnarray}
H_{EtJ}^{(2)} (s_1,s_2)&=&- t \sum_{j,\sigma} (\tilde
c_{j,\sigma}^\dagger \tilde c_{j+1,\sigma}+h.c.)+ (s_1+s_2) t
\sum_j \tilde n_j \tilde n_{j+1}\label{etj_int}\cr H_{EtJ}^{(3)}
(s_1,s_2) &=& - t \sum_{j,\sigma} (\tilde c_{j,\sigma}^\dagger
\tilde c_{j+1,\sigma}+h.c.)+ 2 s_1 t \sum_j
(\vec{{\mathcal{S}}}_{j} \cdot \vec{{\mathcal{S}}}_{j+1}- {1\over
4}\tilde n_j \tilde n_{j+1}) +(s_2+s_1) t \sum_j \tilde n_j \tilde
n_{j+1} \quad ,\cr
\end{eqnarray}
\begin{multicols}{2}
plus conserved quantities. Here the first group gives
generalizations of the infinite $U$ Hubbard model, the latter
being obtained for $s_1=-s_2$; whereas the second group
generalizes the $t-J$ Hamiltonian, which is in fact given by the
choice $s_1=1$, $s_2=-1$.\\Also, models in the second group all
have algebraic structures of type $F^{3-l}B^l$ (these are four, in
one to one correspondence with the four possible choices of signs
$s_1$ and $s_2$); whereas models in the first group are of type
$F^{2-l}B^l$. In particular, infinite $U$ Hubbard model is of type
$FB$ (as expected from the known spinless fermion-like spectrum),
and $t-J$ model turns out to be of type $F^3$. \\Interestingly,
while integrability of models of the second group has already been
discussed in \cite{tJ}, apparently 2 ($V=\pm 2 t$) out of the four
models in the first group have not been studied elsewhere from
this point of view.\\Among the models of first and second type the
choices of $s_1=s_2$ are physically interesting, describing a
model of interacting electrons in the limit of large on-site
Coulomb repulsion and finite ($=\pm 2 t$) neighboring sites
Coulomb interaction. In the first group, for instance, the
repulsive case ($s_1=s_2=1$) corresponds to a $F^2$ model (the
ground state and excited states of which being nicely discussed in
\cite{SUT}), whereas the attractive case is a $B^2$ model. Both
should be compared with the plane infinite $U$ Hubbard case (no
Coulomb repulsion between neighboring sites), in order to
understand whether neighboring sites Coulomb interaction can
induce a quantum phase transition.

\section{Conclusions}
\label{secVIII} In this paper we reviewed a recently developed
method to determine integrable electron models, which amounts in
finding solutions of the YBE of polynomial form. The method was
applied to the simpler case of first degree polynomial, and was
solved in full generality in this case (for any dimension of the
on-site vector space); solutions are all and only generalized
permutators. These were classified in terms of Sutherland species,
which allow for a complete derivation of the spectrum in the case
of models of $FB$ type, as well as for the calculation of the
ground state energy in many other cases. We explicitly applied the
scheme to the classification and solution of integrable electron
models, both in the constrained and in the unconstrained case
($d_V=3,4$ respectively).\\To conclude, it must be stressed that
in fact the scheme developed just requires to have an Hamiltonian
which --when represented in matrix form on two neighboring sites--
has the same structure of a generalized permutator. Hence we
expect the method to be capable of generating integrable models
also in physical context different from correlated electrons. For
instance, spin $S$ Hamiltonians have a local vector space of
dimension $d_V=2 S+1$, implying that models discussed in terms of
constrained fermions in the previous section could also have an
interpretation as spin 1 Hamiltonian. Work is in progress along
these lines.

\end{multicols}

\begin{references}
\bibitem{nota1} In the case of $s$-orbital
electrons we thus have $h^{(1)}_j=c^{\dagger}_{j,\uparrow} \, ; \,
h^{(2)}_j=c^{\dagger}_{j,\downarrow} \, ; \, h^{(3)}_j=1 \, ; \,
h^{(4)}_j = c^{\dagger}_{j,\downarrow} c^{\dagger}_{j,\uparrow}$
\bibitem{nota2} Due to the order defined by eq.(\ref{vectors-s})
we have $p(1)=p(2)=1$ and $p(3)=p(4)=0$.
\end{references}
\end{document}